\documentclass[useAMS,usenatbib]{mn2e}
\usepackage{amsmath}
\usepackage{natbib}
\usepackage{graphicx}
\usepackage{color}
\usepackage{rotating}
\usepackage{nicefrac}
\usepackage{txfonts}
\usepackage{subfigure}
\usepackage{longtable}
\usepackage[percent]{overpic}
\usepackage{xspace}
\usepackage{url}
\usepackage{scalefnt}
\usepackage{amssymb}
\usepackage[percent]{overpic}
\usepackage{multirow,tabularx}

\usepackage{epsfig}
\usepackage{amsmath}
\usepackage{rotating}
\usepackage{natbib}
\usepackage{enumerate}
\usepackage{multirow}

\usepackage{array}
\usepackage{appendix}
\usepackage{comment}

\def \aap{A\&A}

\def \apjl{ApJ}
\def \apjs{ApJS}

\def\gsim{\;\rlap{\lower 2.5pt
 \hbox{$\sim$}}\raise 1.5pt\hbox{$>$}\;}
\def\lsim{\;\rlap{\lower 2.5pt
   \hbox{$\sim$}}\raise 1.5pt\hbox{$<$}\;}

\def\ias15{{\scalefont{0.9}\texttt{IAS15}}\xspace}
\def\rebound{{\scalefont{0.9}\texttt{REBOUND}}\xspace}
\def\emcee{{\scalefont{0.9}\texttt{EMCEE}}\xspace}

\title{Reanalysis of radial velocity data from the resonant planetary system HD128311}

\author[Hanno Rein]{
Hanno Rein$^{1}$
\\
$^1$University of Toronto, Department of Environmental and Physical Sciences, 1265 Military Trail,
Toronto Ontario M1C 1A4\\
\hspace{0.8mm} e-mail: \url{hanno.rein@utoronto.ca}
\vspace{-5mm}
}

\date{
\vspace{-5mm}
Submitted: 25 November 2014 --- Revised: 13 December 2014 -- Accepted: 22 December 2014}

\begin{document}
\maketitle

%%%%%%%%%%%%%%%%%%%%%%%%%%%%%%%%%%%%%%%%%%%%%
%%%%%%%%%%%%%%%%%%%%%%%%%%%%%%%%%%%%%%%%%%%%%
%%%%%%%%%%%%%%%%%%%%%%%%%%%%%%%%%%%%%%%%%%%%%
\begin{abstract}
The multi-planetary system HD128311 hosts at least two planets.
Its dynamical formation history has been studied extensively in the literature. 
We reanalyse the latest radial velocity data for this system with the affine-invariant Markov chain Monte Carlo sampler \emcee.
Using the high order integrator \ias15, we perform a fully dynamical fit, allowing the planets to interact during the sampling process.
A stability analysis using the MEGNO indicator reveals that the system is located in a stable island of the parameter space.
In contrast to a previous study, we find that the system is locked in a~2:1 mean motion resonance.
The resonant angle $\varphi_1$ is librating with a libration amplitude of approximately~$37$${}^\circ$.
The existence of mean motion resonances has important implication for planet formation theories.
Our results confirm predictions of models involving planet migration and stochastic forces.
\end{abstract}

\begin{keywords}
methods: numerical --- gravitation --- planets and satellites: dynamical evolution and stability --- stars: individual (HD 128311)
\end{keywords}

%%%%%%%%%%%%%%%%%%%%%%%%%%%%%%%%%%%%%%%%%%%%%
%%%%%%%%%%%%%%%%%%%%%%%%%%%%%%%%%%%%%%%%%%%%%
%%%%%%%%%%%%%%%%%%%%%%%%%%%%%%%%%%%%%%%%%%%%%
\section{The planetary system HD128311}
The first planet around HD128311 was discovered by \cite{Butler2003}.
A second planet was found two years later \citep{Vogt2005}.
Both planets are most likely gas giants with minimum masses of $1.8~{\rm M}_{\rm jup}$ and $3.2~{\rm M}_{\rm jup}$.
A third signal has been discovered by \cite{McArthur2014} but its planetary nature has yet to be confirmed.
Soon after its discovery HD128311 began to emerge as a test bed of planet formation scenarios within the scientific community. 
A large number of groups studied the formation and evolution of this system with particular focus on the system's dynamical history
\citep{Beauge2006,Quillen2006,SandorKley2006,Sandor2007,Michtchenko2008,Meschiari2009,Crida2008,LeeThommes2009,Voyatzis2009,BarnesGreenberg2006,Raymond2008,ReinPapaloizou09,Lecoanet2009,GayonMarktBois2009,GozdziewskiKnacki2006,Erdi2007,Giuppone2010}.

The most important dynamical feature of HD128311 is its proximity to a 2:1 mean motion resonance (MMR). 
Early observations supported the idea that HD128311 is in resonance, although some of the orbital solutions were dynamically unstable on short timescales \citep{Vogt2005}.
The most recent work on HD128311 by \cite{McArthur2014} includes new radial velocity data, a recalibration of older data sets and astrometric and photometric constraints.
The authors found that the planets in their best fit model are not in a MMR.
Whether this system is in a MMR or not is an important constraint for planet formation scenarios.
For example a system in a MMR favours the idea that giant planets migrated when they were still embedded in a protoplanetary disc \citep{LeePeale2002,ReinPapaloizou09}.

In this letter we reanalyse the combined radial velocity (RV) datasets including the recalibrated data from the Hobby-Eberly Telescope (HET) and the Lick Observatory.
Our RV data is thereby equivalent to that of \cite{McArthur2014} but we do not take into account the astrometric observations in our model which only constrain the system's inclination.
In contrast to previous studies, we use a fully dynamical model, allowing planets to interact, rather than being on fixed Keplerian orbits.
We use a use a modern Markov chain Monte Carlo sampler to explore the high dimensional parameters space and converge on a new set of orbital parameters.
The details are described in Section~\ref{sec:mcmc}.
We then perform long term orbit integrations to study the stability in Section~\ref{sec:longterm} using the fast chaos indictor MEGNO \citep{CincottaSimo2000}.
In Section~\ref{sec:resonance} we conclude by discussing the resonant nature of the system and the implications for planet formation scenarios involving stochastic migration.

%%%%%%%%%%%%%%%%%%%%%%%%%%%%%%%%%%%%%%%%%%%%%
\section{Markov chain Monte Carlo}\label{sec:mcmc}

\begin{figure*}
 \centering \resizebox{0.99\textwidth}{!}{\includegraphics[trim=2cm 0 3cm 0]{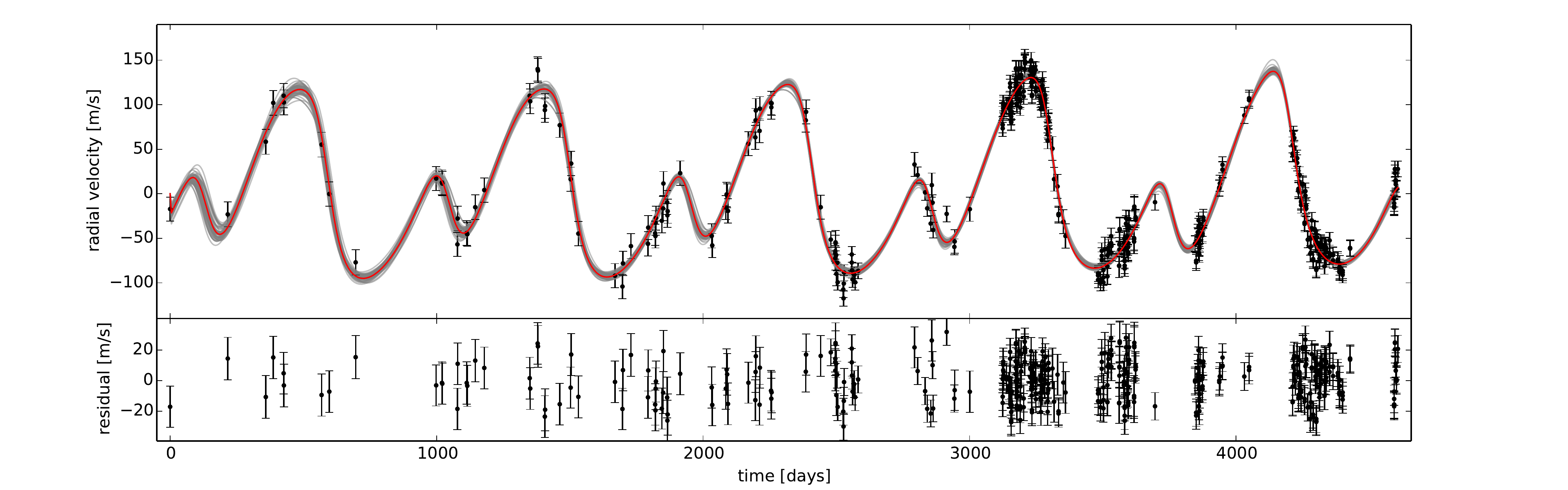}}
 \caption{
Radial velocity data and model.
The time is relative to JD2450983.83.
Top panel: RV data points, MCMC mean (red), MCMC samples (gray). 
Bottom panel: residual errors not including jitter $s$. 
\label{fig:rv}}
\end{figure*}

We use a Markov chain Monte Carlo method to fit the observed radial velocity datasets.
In comparison to the fit performed by \cite{McArthur2014} we use a fully dynamical two planet model.
For this purpose, we couple the high order \ias15 \citep{ReinSpiegel2014} integrator which is available within the \rebound package \citep{ReinLiu2012} to the \emcee package.
\emcee is an open-source, parallelized affine-invariant Marcov chain Monte Carlo sampler written in python \citep{emcee}.

We assume co-planarity of the planets but allow the orbital plane to be inclined from the line of sight by an angle $i$.
The central object has a fixed mass of $M_* = 0.828 M_\odot$.
We have four orbital elements and one mass parameter, $m\sin(i)$, for each planet.
The orbital parameters are the period $P$, the eccentricity $e$, the argument of periapsis $\omega$ and the mean anomaly $M$.
For the sampling process, we perform a coordinate transformation to $h=e\sin(\omega)$ and $k=e\cos(\omega)$ variables.
This allows us to avoid the coordinate singularity at $e=0$ and speed up convergence \citep[for a discussion see][]{Hou2012}.
We use a Jacobi coordinates, i.e. the outer planet's orbital parameters are given with respect to the centre of mass of the star and the inner planet.
Building up on the work of \cite{McArthur2014}, we further include an individual offset $\gamma$ and jitter parameter $s$ for each RV dataset (four for Lick, one for HET).
This allows the model to account for variations in the offsets for different instruments as well as variations in the Lick detectors between observing runs.
The total number of free parameters is thus~21.

A total of 500 walkers are evolved for several thousand generations.
Both angles $\omega$ and $M$ as well as the eccentricity $e$ for each planet have a uniform priors. 
We follow \cite{Hou2012} and use uninformative Jefferys priors for the period, the planet mass as well as the jitter parameters.
We initialize the walkers with mass and period parameters that are roughly those of previous results to speed up convergence.
Our experiments showed that the precise details on how the walkers are initialized do not matter.
Finally, we perform a simple declustering algorithm after 1000 generations by removing those samples from the ensemble that have a likelihood significantly smaller than the best samples \citep{Hou2012}.

After the MCMC has converged and has been declustered, we sample it over 1250~generations and calculate the mean of all parameters as well as the two sigma confidence intervals.
The results are listed in Table~\ref{tab:parameters}.
In Figure~\ref{fig:rv} we plot the observed radial velocity datapoints with the offset adjusted according to our model. 
The red line corresponds to the mean solution of Table~\ref{tab:parameters}.
The gray lines correspond to 50 randomly drawn samples from the MCMC posterior distribution.

%We list the priors for each parameter in Table~\ref{tab:prior}. 
%Note that the priors are uninformative in they sense that they do not include current knowledge of planet occurrence rates and system architectures.

%\begin{table}
%\caption{Priors used in the MCMC. \label{tab:prior}}
% \begin{tabular}{lll}
% \hline\hline Parameter && Prior   \\\hline
% inclination&$i$  & uniform on $[0,\pi/2]$\\  
% eccentricity&$e$  & uniform on $[0,1]$\\  
% argument of periaspis&$\omega$  & uniform on $[0,2\pi]$\\  
% mean anomaly&$M$  & uniform on $[0,2\pi]$\\  
% minimum mass&$m$  & Jefferys, $\log \frac{1}{m+m_0}$, $m_0=1 M_{\rm jup}$\\  
% period&$P$  & Jefferys, \\  
% offset&$\gamma$  & Jefferys, \\  
% jitter&$s$  & Jefferys, \\  
% \end{tabular}
%\end{table}

%%%%%%%%%%%%%%%%%%%%%%%%%%%%%%%%%%%%%%%%%%%%%
\section{Long term stability}\label{sec:longterm}
\begin{figure*}
 \centering \resizebox{0.99\textwidth}{!}{\includegraphics[trim=1.2cm 0.5cm 2.2cm 1.0cm]{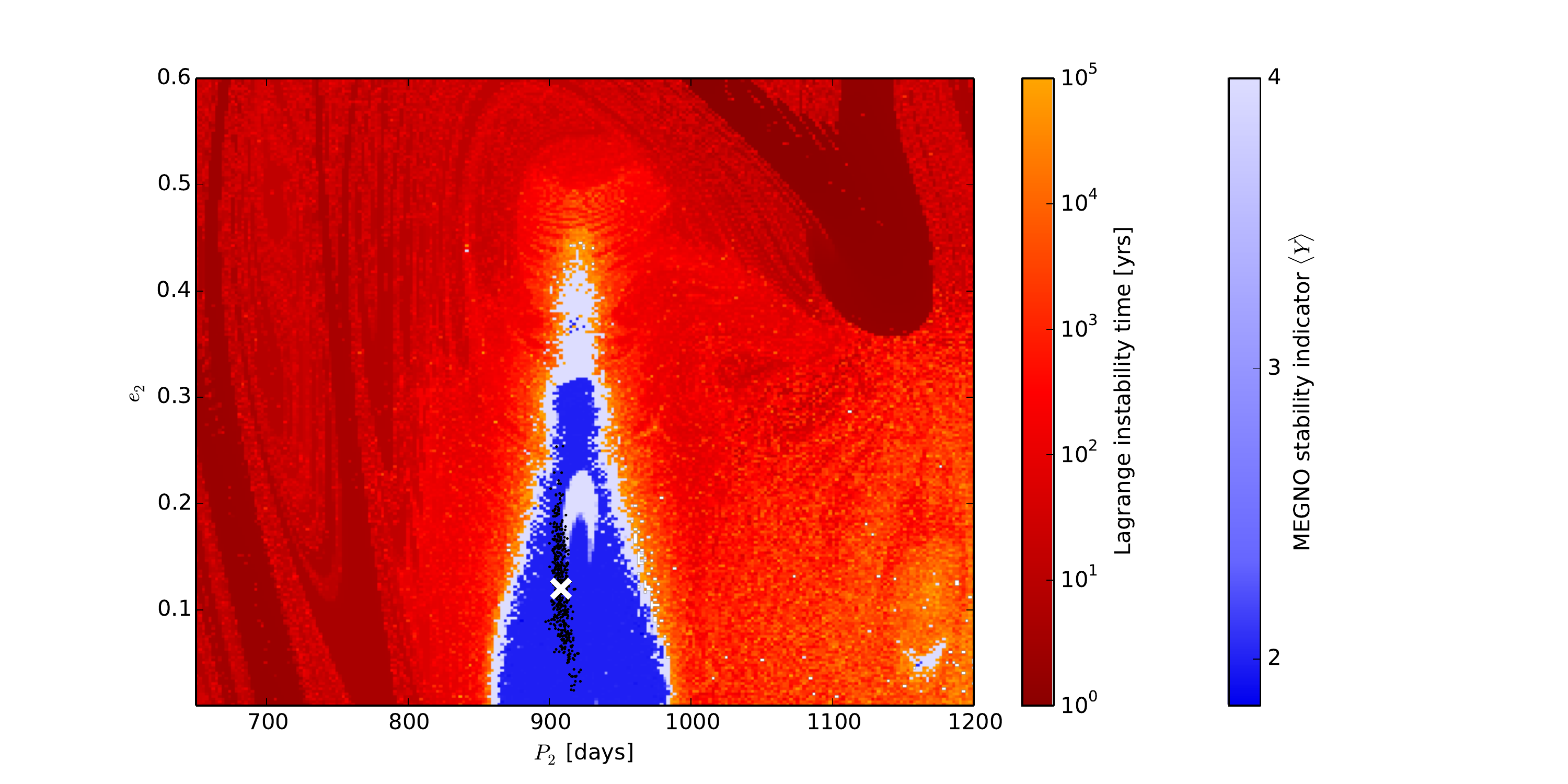}}
 \caption{
Lagrange stability plot in the vicinity of the MCMC solution.
The axes correspond to the period and eccentricity of the outer planet.
Red regions are Lagrange unstable. 
Blue regions are stable for the duration of the integration. 
Dark blue regions are stable quasi-periodic solutions according to the MEGNO indicator.
The white cross shows the nominal mean orbital parameters listed in Table~\ref{tab:parameters}.
The black dots show samples of the MCMC posterior. 
\label{fig:megno}}
\end{figure*}

We focus on parameters close to those of the converged MCMC samples and run a total of $57\,600$ realisations of HD128311 for $t_{\rm{max}}=10^5$~years to map out the structure of the phase space.
A smaller subset was integrated for $t_{\rm max}=10^6$~years, but no qualitative differences could be observed.
As in the previous section we use \rebound and the high order \ias15 integrator for these long term integrations.
We declare a system unstable if at least one of the planets gets ejected, if the planets have a close encounter, or if the semi-major axis of at least one planet changes by more than 50\%.
We refer to this definition as Lagrange stability and call the time until instability Lagrange timescale.

For systems that are Lagrange stable, we measure the Mean Exponential Growth of Nearby Orbits, MEGNO \citep{CincottaSimo2000}.
MEGNO is a fast chaos indicator similar to the traditional Lyapunov exponent, but gives more useful results on shorter timescales.
We compute the MEGNO value $\langle Y \rangle$ by integrating the variational equations \citep{MikkolaInnanen1999} using \ias15.
For a definition of $\langle Y\rangle$ and a detailed discussion of the MEGNO indicator see \cite{CincottaSimo2000} and \cite{Gozdziewski2001}\footnote{Note that there is a typo in the denominator of $\dot \delta$ in Equation 5 of \cite{Gozdziewski2001}.}.

In Figure~\ref{fig:megno} we show a slice of the parameter space in the plane of the outer planet's period and eccentricity, $P_2$ and $e_2$.
All other orbital parameters are those listed in Table~\ref{tab:parameters}.
All simulations in red regions are Lagrange unstable.
The shade of red indicates the Lagrange timescale.
Simulations in blue regions remain stable for the entire integration. 
The shade of blue corresponds to the MEGNO~$\langle Y\rangle$. 
A value of $\langle Y\rangle=2$ (dark blue) indicates a stable quasi-periodic orbit, whereas a value larger than 2 indicates chaotic behaviour.
One can see in the figure that seemingly stable systems which are close to the stability boundary are in fact chaotic and thus might become unstable over longer timescales.

The white cross in Figure~\ref{fig:megno} shows the mean orbital parameters of the MCMC sample.
The black dots show the individual MCMC samples from the posterior distribution.
Both the mean solution and all MCM samples are located well within a stable island.
Note that the eccentricity $e_2$ is not well constrained. 
However, the observed RV data is inconsistent with the assumption of a circular orbit.

The observational data is currently not good enough to constrain the mutual inclination of the two planets.
To test the effects of mutually inclined orbits, we ran an additional $14\,400$ simulations with the same initial conditions as before but perturbed each system with a random mutual inclination of $\sim2^\circ$. 
Our results indicate that the system's stability is not significantly affected by a small amount of mutual inclination over a million year timescale.

%%%%%%%%%%%%%%%%%%%%%%%%%%%%%%%%%%%%%%%%%%%%%
\section{Mean Motion Resonance and conclusions}\label{sec:resonance}
The period ratio in our MCMC sample, $P_2/P_1$, is within 2\% of a 2:1 mean motion resonance.
The period ratio alone is not a sufficient criteria for a MMR.
To test whether the system is in a MMR or not we therefore monitor the two resonant angles $\varphi_{1}=\lambda_1-2\lambda_2+\omega_{1}$ and $\varphi_{2}=\lambda_1-2\lambda_2+\omega_{2}$, where $\lambda$ is the mean longitude.
In \emph{all} of our MCMC samples, $\varphi_1$ is librating around $0{}^\circ$. 
The libration amplitudes range from close to $0{}^\circ$ to up to 90${}^\circ$ with a mean of 37${}^\circ$. 
The resonant angle $\varphi_2$ is librating in the majority but not all of the samples. 
The mean libration amplitude for $\varphi_2$ is 76.5${}^\circ$.
Our findings differ qualitatively from those of \cite{McArthur2014} who found that the planets are not in a MMR.

The existence of a MMR is an important observable in many planet formation scenarios. 
A system in a MMR supports the idea that giant planets migrated into their current position while they were still embedded in a protoplanetary disc \citep{LeePeale2002}.
Furthermore, \cite{ReinPapaloizou09} predicted that if the system did undergo a phase of stochastic migration, then $\varphi_1$ should librate at moderate amplitudes whereas $\varphi_2$ should be close to the separatrix of libration. 
Our results are in perfect agreement with this prediction, thus supporting the idea that both planet migration and stochastic forces occurred during the system's evolution.

%The HD128311 system continues to be an excellent test best for studying planet formation scenarios.

\section*{Acknowledgments}
We thank Barbara McArthur for sharing the latest radial velocity dataset.
Daniel Tamayo and an anonymous referee provided valuable feedback.
This research has been supported by the NSERC Discovery Grant RGPIN-2014-04553.

\begin{table}
\caption{Model parameters from MCMC sample. \label{tab:parameters}}
\def\arraystretch{1.4}
\begin{tabular}{lll}\hline\hline Parameter && Value and 2-$\sigma$ confidence interval   \\
\hline
epoch  & JD &  2450983.83 (fixed) \\
stellar mass & M${}_*$ &  $0.828$  M${}_\odot$ (fixed) \\
inclination & $i$ &  $63.8^{+23.7}_{-35.9}$  ${}^\circ$ \\
\hline \multicolumn{3}{c}{Planet 1}\\\hline
minimum mass & $m_1\sin(i)$  & $1.83^{+0.15}_{-0.18}$   M${}_{\rm jup}$\\
period & $P_1$ & $460.1^{+4.2}_{-3.6}$   $\rm{days}$\\
eccentricity & $e_1$ & $0.30^{+0.03}_{-0.04}$  \\
argument of periapsis&$\omega_1$  & $-76.2^{+6.4}_{-9.2}$  ${}^\circ$ \\
mean anomaly&$M_1$  & $259.2^{+11.9}_{-12.6}$   ${}^\circ$\\
\hline \multicolumn{3}{c}{Planet 2}\\\hline
minimum mass & $m_2\sin(i)$  & $3.20^{+0.08}_{-0.08}$   M${}_{\rm jup}$\\
period & $P_2$ & $910.7^{+7.6}_{-6.0}$   $\rm{days}$\\
eccentricity & $e_2$ & $0.12^{+0.08}_{-0.06}$  \\
argument of periapsis&$\omega_2$  & $-19.7^{+23.2}_{-12.0}$  ${}^\circ$ \\
mean anomaly&$M_2$  & $184.2^{+20.0}_{-10.7}$   ${}^\circ$\\
\hline \multicolumn{3}{c}{Lick Radial Velocities starting JD 2450983}\\\hline
offset${}^{\rm{a}}$&$\gamma_1$ &  $1.2^{+4.1}_{-3.6}$  $\rm{m/s}$ \\
jitter${}^{\rm{b}}$ &$s_1$ &  $1.5^{+3.8}_{-1.4}$  $\rm{m/s}$ \\
\hline \multicolumn{3}{c}{Lick Radial Velocities starting JD 2451409}\\\hline
offset${}^{\rm{a}}$&$\gamma_2$ &  $7.9^{+11.4}_{-13.0}$  $\rm{m/s}$ \\
jitter${}^{\rm{b}}$ &$s_2$ &  $5.2^{+10.9}_{-4.9}$  $\rm{m/s}$ \\
\hline \multicolumn{3}{c}{Lick Radial Velocities starting JD 2452333}\\\hline
offset${}^{\rm{a}}$&$\gamma_3$ &  $11.2^{+21.0}_{-17.6}$  $\rm{m/s}$ \\
jitter${}^{\rm{b}}$ &$s_3$ &  $6.4^{+17.1}_{-5.7}$  $\rm{m/s}$ \\
\hline \multicolumn{3}{c}{Lick Radial Velocities starting JD 2452515}\\\hline
offset${}^{\rm{a}}$&$\gamma_4$ &  $3.3^{+10.4}_{-10.0}$  $\rm{m/s}$ \\
jitter${}^{\rm{b}}$ &$s_4$ &  $2.8^{+6.6}_{-2.6}$  $\rm{m/s}$ \\
\hline \multicolumn{3}{c}{HET Relative Radial Velocities}\\\hline
offset${}^{\rm{a}}$&$\gamma_5$ &  $0.6^{+1.5}_{-1.4}$  $\rm{m/s}$ \\
jitter${}^{\rm{b}}$ &$s_5$ &  $2.2^{+1.9}_{-2.0}$  $\rm{m/s}$ \\
\end{tabular}

\vspace{2mm}
\textbf{Notes:}\\
${}^{\rm a}$ Offsets are relative to those found by \cite{McArthur2014} and are all consistent with zero.\\
${}^{\rm b}$ Jitter is added on top of the reported RV errors which do not include stellar variability and additional instrumental noise.
\vspace{20mm}
\end{table}

%`\vfill\eject
%\onecolumn
\bibliography{full}

\end{document}